\documentclass[10pt,conference]{IEEEtran}
\IEEEoverridecommandlockouts

\usepackage[]{collab}
\usepackage{graphicx}
\usepackage{subfigure}
\usepackage{balance}
\usepackage{amsfonts}
\usepackage{enumitem}
\collabAuthor{zb}{teal}{Zhuangbin Chen}

\usepackage{amsmath}
\newcommand\numberthis{\addtocounter{equation}{1}\tag{\theequation}}

\newcommand{\nm}{CMAnomaly\xspace}
\newcommand{\cpn}{Huawei Cloud\xspace}

\begin{document}

\title{Practical Anomaly Detection over Multivariate Monitoring Metrics for Online Services (PER)}



\author{\large 
Jinyang Liu\IEEEauthorrefmark{2}, 
Tianyi Yang\IEEEauthorrefmark{2},  
Zhuangbin Chen\IEEEauthorrefmark{3}$^{**}$\thanks{\hspace{-2ex}$^{**}$Zhuangbin Chen is the corresponding author.}, 
Yuxin Su\IEEEauthorrefmark{3}, \\ 
Cong Feng\IEEEauthorrefmark{4}, Zengyin Yang\IEEEauthorrefmark{4}, Michael R. Lyu\IEEEauthorrefmark{2}\\
\IEEEauthorblockA{
\IEEEauthorrefmark{2}The Chinese University of Hong Kong, Hong Kong SAR, China, \{jyliu, tyyang, lyu\}@cse.cuhk.edu.hk\\
\IEEEauthorrefmark{3}School of Software Engineering, Sun Yat-sen University, Zhuhai, China, \{chenzhb36,suyx35\}@mail.sysu.edu.cn\\
\IEEEauthorrefmark{4}Computing and Networking Innovation Lab, Huawei Cloud Computing Technology Co., Ltd, China,\\ \{fengcong5, yangzengyin\}@huawei.com
}
}

\maketitle

\begin{abstract}

As modern software systems continue to grow in terms of complexity and volume, anomaly detection on multivariate monitoring metrics, which profile systems' health status, becomes more and more critical and challenging. In particular, the dependency between different metrics and their historical patterns plays a critical role in pursuing prompt and accurate anomaly detection. 
Existing approaches fall short of industrial needs for being unable to capture such information efficiently.
To fill this significant gap, in this paper, we propose CMAnomaly, an anomaly detection framework on multivariate monitoring metrics based on collaborative machine. The proposed collaborative machine is a mechanism to capture the pairwise interactions along with feature and temporal dimensions with linear time complexity. Cost-effective models can then be employed to leverage both the dependency between monitoring metrics and their historical patterns for anomaly detection. The proposed framework is extensively evaluated with both public data and industrial data collected from a large-scale online service system of Huawei Cloud.
The experimental results demonstrate that compared with state-of-the-art baseline models, CMAnomaly achieves an average F1 score of 0.9494, outperforming baselines by $6.77\%\sim10.68\%$, and runs $10\times\sim 20\times$ faster. Furthermore, we also share our experience of deploying \nm in Huawei Cloud.



\end{abstract}



\maketitle

\section{Introduction}

As modern software systems, e.g., online service systems, have grown to an unprecedented scale, failures become inevitable, leading to significant revenue loss and user dissatisfaction~\cite{DBLP:conf/sigsoft/ChenKLZZXZYSXDG20,DBLP:conf/usenix/ZhangLXQ0QDYCCW19,DBLP:conf/icse/DangLH19,DBLP:journals/corr/abs-2009-07237,DBLP:conf/sigsoft/HeLLZLZ18}. To understand the health status of an online service system, various monitoring metrics such as network traffic, server response delay, and CPU usage rate are closely monitored~\cite{DBLP:journals/corr/abs-2302-05092, chen2022adaptive, liu2023incident, liu2023prism, li2022intelligent}. Anomaly detection over the monitoring metrics is a crucial approach to ensure the reliability and availability of the system, which aims to discover unexpected events or rare items in data.
Many efforts~\cite{DBLP:conf/kdd/RenXWYHKXYTZ19/microsoft, DBLP:conf/www/XuCZLBLLZPFCWQ18/donut, DBLP:journals/corr/abs-2004-00433/univariate_survey, DBLP:journals/tkde/GuptaGAH14/anomaly_survey2} have been devoted to univariate metric anomaly detection, which deals with only one single type of metric. However, more often than not, different types of monitoring metrics collectively can indicate the occurrence of anomalies more precisely~\cite{DBLP:conf/usenix/ZhangLXQ0QDYCCW19}. Different system components (e.g., microservices for different functionalities) are tightly coupled, and failures tend to trigger anomalies in multiple monitoring metrics. For example, a problematic load balance server is often accompanied by a burst on both round-trip delay and in-bound traffic rate, which will further increase CPU utilization. However, a sudden rise in CPU utilization alone could be caused by regular service upgrades, which should not be regarded as an anomaly. Compared to univariate metric, anomaly detection on multivariate monitoring metrics is more challenging. We have identified three main reasons:

\textbf{Demanding industrial requirements}. In modern software systems, minutes of downtime could cause an expensive drain on company revenue. Therefore, suspicious anomalies should be quickly identified. Moreover, as systems continuously undergo feature upgrades and system renewal, the patterns of multivariate monitoring metrics may shift accordingly. To accommodate such ever-changing pattern shifts, the anomaly detection model must support fast model retraining.

\textbf{Large-volume and noisy data}. Although terabytes and even petabytes of metric data are being generated everyday, most of them do not contain much valuable information. For example, a significant portion of monitoring metrics only records plain system runtime behaviors. Moreover, a low signal-to-noise ratio is also a critical issue. Thus, multivariate metric anomaly detection with the presence of noise is challenging.

\textbf{Sparsity of monitoring metrics' dependency}. In industrial systems, hundreds or even thousands of monitoring metrics are being monitored. The dependencies among monitoring metrics are very sparse, i.e., most monitoring metrics are not or weakly dependent on other monitoring metrics. Therefore, how to automatically learn the dependencies among different monitoring metrics is critical towards efficient multivariate metric anomaly detection.

In the literature, many studies have shifted to anomaly detection on multivariate monitoring metrics, which mainly resorts to different deep learning-based models. For example, OmniAnomaly~\cite{DBLP:conf/kdd/SuZNLSP19/omni} proposes to learn the normal patterns of multivariate time series by modeling data distribution through stochastic latent variables. Anomalies are then determined by reconstruction probabilities. Similarly, Malhotra et al.~\cite{DBLP:journals/corr/MalhotraRAVAS16} used an LSTM-based (long short-term memory) encoder-decoder network to learn time series's normal patterns and Zhang et al.~\cite{DBLP:conf/aaai/ZhangSCFLCNZCC19} used an attention-based convolutional LSTM network. 
Although tremendous progress has been made, we still observe two major limitations of existing approaches: 1) the interactions among monitoring metrics are not explicitly modeled, and 2) the efficiency falls behind industrial needs. 
Specifically, previous approaches~\cite{DBLP:conf/kdd/SuZNLSP19/omni,DBLP:journals/corr/MalhotraRAVAS16, DBLP:journals/corr/abs-1711-00614/lstm-vae} detect anomalies on multivariate monitoring metrics mainly by stacking different types of monitoring metrics into a feature matrix and feeding it to sophisticated neural network models. In this manner, each metric corresponds to one feature, and monitoring metrics' interactions are implicitly learned at the cost of complicated model design, resulting in inevitable false alarms. More recent studies tackle this problem by constructing an $m\times m$ metric inner-product matrix~\cite{DBLP:conf/aaai/ZhangSCFLCNZCC19} or a complete graph~\cite{DBLP:conf/icdm/ZhaoWDHCTXBTZ20} for $m$ different monitoring metrics to capture the pairwise metric interaction, both of which yield an $\mathcal{O}(m^2)$ computation complexity. Similarly, heavy models such as graph neural networks~\cite{DBLP:journals/tnn/ScarselliGTHM09} are employed to learn monitoring metrics' dependencies by feeding the matrix or training on the graph. Different from them, we argue that by properly modeling the interactions of monitoring metrics along with feature and temporal dimensions, cost-effective neural network models can be leveraged for anomaly detection.

In this paper, to overcome the aforementioned limitations, we propose CMAnomaly, an efficient unsupervised model for anomaly detection over multivariate monitoring metrics. Specifically, we propose a collaborative machine to conduct cross-feature and cross-time metric interactions by taking the inner product of pairwise monitoring metrics and pairwise timestamps. Moreover, to capture the existing sparse dependency, we assign a weight to each metric (or timestamp), and the importance of a pair of dependencies is measured by the product of corresponding weights. However, such a weighted interactions mechanism still costs quadratic computation as previous approaches do. To alleviate this problem, inspired by factorization machines~\cite{DBLP:conf/icdm/Rendle10/FM}, we reformulate the proposed collaborative machine and reduce the computation from quadratic time complexity to linear complexity, which ensure the efficiency of CMAnomaly.  After that, to consider both the feature and temporal factors simultaneously, the results of interactions from both aspects are concatenated and fed to a cost-effective multi-layer perceptron (MLP). Particularly, the model is trained by learning from the historical normal patterns of monitoring metrics and predicting their future states. Anomalies are then identified based on prediction errors. Such learning paradigms have been proven to be effective in many related studies~\cite{DBLP:conf/kdd/HundmanCLCS18/lstm_ndt, DBLP:conf/icdm/ZhaoWDHCTXBTZ20}.
To sum up, in this paper, we make the following major contributions:
\begin{itemize}[leftmargin=*, topsep=0pt]
    \item We propose CMAnomaly, an efficient unsupervised model for anomaly detection over multivariate monitoring metrics (\S~\ref{sec:methodology}). CMAnomaly explicitly models the interactions among monitoring metrics along both temporal and feature dimensions. More importantly, we reduce the computation of our model from quadratic to linear complexity. In doing this, anomaly detection can be performed efficiently to process a large volume of data in the real world.

    \item We conduct extensive experiments with both public and industrial data collected from a large-scale online service provider, Huawei Cloud (\S~\ref{sec:experiments}). The experimental results demonstrate that CMAnomaly achieves 0.9494 F1 score with $4.45\%$ improvement over the state-of-the-art approaches. Moreover, CMAnomaly can run $10\times \sim 20\times$ faster than the baseline models.
    
    \item We have successfully incorporated CMAnomaly into the troubleshooting system of the Huawei Cloud. Feedback from on-site engineers confirms its practical benefits conveyed to various online services and products. We also share the hands-on experience story to benefit the community (\S~\ref{sec:discussions}).
\end{itemize}

\section{Background}
\label{sec:background}

\subsection{Dependency of Multivariate Monitoring Metrics}

In recent years, the complexity and scale of modern software systems are growing at a rapid speed. Various types of monitoring metrics that profile an entity's health status need to be closely monitored~\cite{DBLP:conf/kdd/SuZNLSP19/omni}. Particularly, an entity can be a logical one like a service or a physical one like a computing server. They provide system operators and engineers with the most prompt and direct way to understand the system. As different system components are tightly-coupled and work collaboratively, there exists some dependency among monitoring metrics, i.e., Multivariate monitoring metrics. Consequently, when failures are encountered, the dependent monitoring metrics are likely to exhibit abnormal behaviors simultaneously. Such dependencies reflect some intrinsic properties of the system and thus provide more reliable evidence for multivariate monitoring metric anomaly detection.

\begin{figure}[h]
\centering
\includegraphics[scale=0.38]{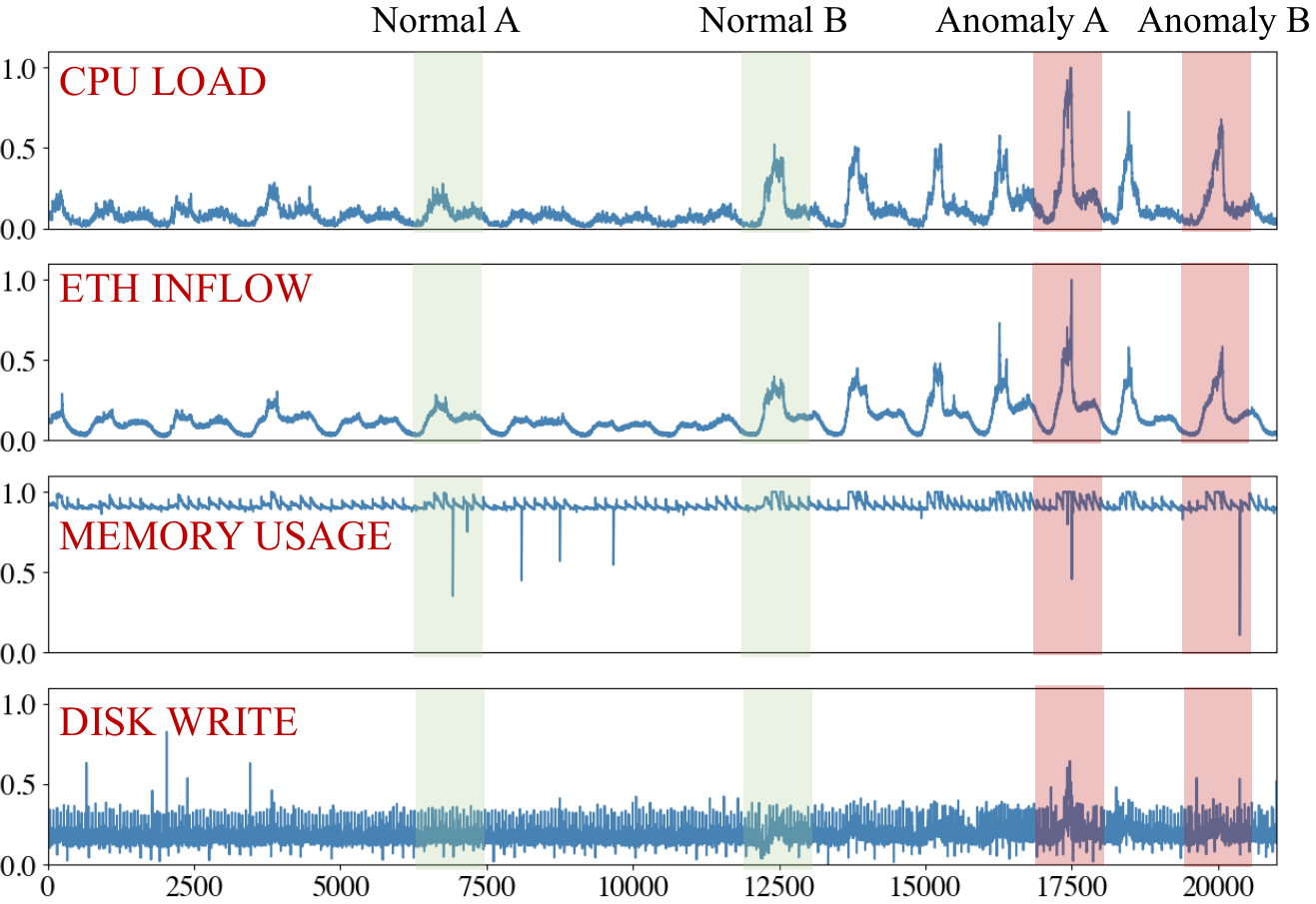}
\caption{Multivariate monitoring metrics Snippet from Server Machine Dataset}
\label{fig: dependency}
\end{figure}

A real-world example is provided in Figure~\ref{fig: dependency}, which is from a public dataset released by~\cite{DBLP:conf/kdd/SuZNLSP19/omni}. The four monitoring metrics in the figure constitute a typical set of dependent monitoring metrics. Particularly, \textit{CPU LOAD} and \textit{ETH INFLOW} are strongly correlated as their curves exhibit a highly similar trend. Without considering such dependency, we will miss a complete picture of the system's health status. For example, in the segment marked as \textit{Normal A}, we can see a clear spike in \textit{MEMORY USAGE}, which would be flagged as an anomaly without a glance at the other three monitoring metrics. Similar situation happens to segment \textit{Normal B}, where boots can be clearly seen in both \textit{CPU LOAD} and \textit{ETH INFLOW}. Therefore, we need to consider the full set of multivariate monitoring metrics to pursue an accurate anomaly detection, as shown in segment \textit{Anomaly A} and \textit{Anomaly B}. Besides the dependency between monitoring metrics, we can also leverage historical monitoring metric patterns to reduce false positives. Specifically, in Figure~\ref{fig: dependency}, all monitoring metrics have witnessed some abnormal spikes in history. However, they do not necessarily indicate the occurrence of failures. By learning such historical patterns, our model will not be too sensitive to spikes and will be able to distinguish benign change patterns from anomalous ones.


\subsection{Problem Statement}

Anomaly detection on system monitoring metrics is crucial for proactive and prompt troubleshooting, aiming to discover abnormal status exhibited by the monitoring metrics. Such anomalies often indicate that a system encounters faults or attacks. In this paper, we focus on anomaly detection for an entity based on the multivariate monitoring metrics collected from it at equal-space timestamps~\cite{DBLP:conf/kdd/SuZNLSP19/omni}. The problem can be formally defined as follows.

The input of multivariate monitoring metrics can be represented as $X\in \mathbb{R}^{n\times m}$, where $n$ is the number of different monitoring metrics and $m$ is the number of observations. The $t^{th}$ row of $X$, denoted as $x_t=[x_t^1, x_t^2, ..., x_t^m]$, is a m-dimensional vector containing the observation of each monitoring metric at timestamp $t$. Similarly, the $k^{th}$ column of $X$, denoted as $x^k = [x_1^k, x_2^k, ..., x_n^k]$, is a n-dimensional vector containing the observations of the $k^{th}$ monitoring metric. Particularly, we denote $x^k_{i:j} = [x_i^k, x_{i+1}^k, ..., x_j^k]$ as a consecutive sequence of $x^k$ from timestamp $i$ to $j$.

The objective of anomaly detection for multivariate monitoring metrics is to determine whether a given $x_t$ is anomalous, i.e., whether the entity is in abnormal status at timestamp $t$. For each timestamp $t$, our model calculates an anomaly score $s_t\in [0, 1]$, which represents the probability of $x_t$ being anomalous. If $s_t$ is larger than a pre-defined threshold $\theta$, $x_t$ will be predicted as an anomaly. The ground truth $\textbf{y}\in \mathbb{R}^n$ is an n-dimensional vector consisting 0 and 1, where 0 indicates a normal point, and 1 indicates an anomalous one.


\begin{figure*}[!ht]
\centering
\includegraphics[width=0.9\textwidth]{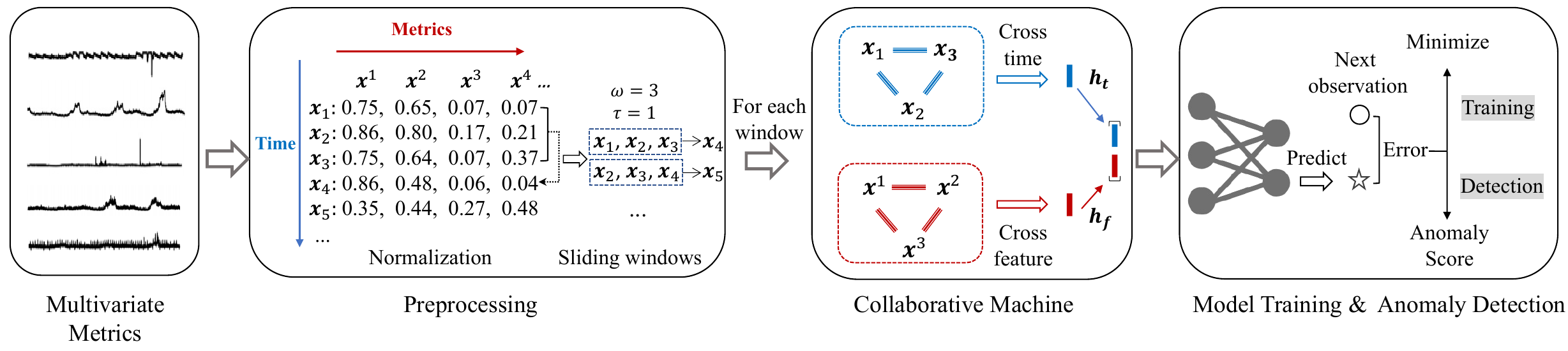}
\vspace{-10pt}
\caption{Overall Framework of CMAnomaly} 
\vspace{-20pt}
\label{fig:overall_framework}
\end{figure*}

\section{Methodology}
\label{sec:methodology}

In this section, we introduce CMAnomaly, our anomaly detection framework for multivariate monitoring metrics. We will begin with an overview of CMAnomaly, followed by a detailed explanation.

\subsection{Overview of CMAnomaly}

In today's software systems, e.g., online service systems, engineers face an overwhelming number of monitoring metrics, which are being generated in a 24×7 non-stop basis. It would be labor-intensive and error-prone to manually investigate each metric for failure detection and diagnosis. CMAnomaly facilitates this process by providing automated and effective multivariate metric anomaly detection.

The overall framework of CMAnomaly is illustrated in Figure~\ref{fig:overall_framework}, which consists of four phases, i.e., \textit{data preprocessing}, \textit{collaborative machine}, \textit{model training}, and \textit{anomaly detection}. The first phase preprocesses the data by applying normalization and window sliding. Particularly, the input types of monitoring metrics can vary depending on the application scenario. These monitoring metrics can be unified to the same range by normalization.
In the subsequent phase, the preprocessed data is input into the collaborative machine, CMAnomaly's core component, which effectively captures interactions among multivariate monitoring metrics across both feature and temporal dimensions. Next, we train a forecasting-based anomaly detection model~\cite{DBLP:conf/kdd/HundmanCLCS18/lstm_ndt,DBLP:conf/ccs/Du0ZS17/deeplog} that identifies anomalies via prediction errors. The collaborative machine enables the use of a cost-effective model architecture, with both phases crucial for a highly accurate and efficient anomaly detector. Finally, the trained model is used to detect anomalies in new observations.

\subsection{Preprocessing}

Different monitoring metrics may have distinct scales, for example, the metric monitoring the CPU execution, i.e., \textit{CPU~USAGE}, is in the range of $0\%$ to $100\%$, However, the metric monitoring the network traffic, i.e., \textit{INBOUND~PACKAGE~RATE} can range from zero to millions of kilobytes. Therefore, data normalization is performed for each individual metric to ensure the robustness of our model. After that, we conduct a sliding window to generate input to the model. Each window preserves the local pattern of multivariate monitoring metrics.

\subsubsection{Data Normalization}

We apply max-min normalization to each individual metric, i.e., $x^k$, as follows:

\begin{equation}
    x^k_{norm} = \frac{x^k - \min(x^k)}{\max(x^k) - \min(x^k)},
\end{equation}

\noindent where the values of $max(x^k)$ and $min(x^k)$ are computed in the training data, which will then be used for test data normalization. For simplicity, we omit the "norm" subscript in the following elaboration.

\subsubsection{Sliding Window}


The sliding window is to partition monitoring metrics along the temporal dimension. Particularly, it consists of two attributes, i.e., window size $\omega$ and stride $\tau$. The stride indicates the forwarding distance of the window along the time axis to generate multivariate metric windows. As the stride is often smaller than the window size, there exists overlapping between two consecutive windows. We denote the $s^{th}$ sliding window as:

\begin{equation}
    X_s = [x_{s\tau}, x_{s\tau + 1}, ..., x_{s\tau + \omega - 1}]
\end{equation}

\noindent where $s\in[0, 1, 2,...]$. $X_s$ together with the observations at the next timestamp of the window, i.e., $\hat{x}_s=x_{s\tau + \omega}$, constitute a pair $(X_s, \hat{x}_s)$, where $X_s\in \mathbb{R}^{\omega\times m}$ and $\hat{x}_s \in \mathbb{R}^m$.




\subsection{Collaborative Machine}

As depicted in Figure~\ref{fig: dependency}, accurate anomaly detection on multivariate monitoring metrics requires consideration of metric dependencies. Historical patterns also offer crucial anomaly detection clues, especially if historical spikes have not resulted in reported failures. By learning these benign change patterns, our model can effectively reduce noise in monitoring metrics.

In this section, we propose the collaborative machine, a mechanism to efficiently capture the interactions of multivariate monitoring metrics along with both the feature and temporal dimensions. Specifically, we allow cross-feature and cross-timestamp interactions by taking the inner-product of pairwise monitoring metrics and pairwise timestamps in a multivariate metric window as shown in Figure~\ref{fig:overall_framework}. The results are two types of vectors, i.e., metric feature vector and temporal vector, which captures the dependencies and historical patterns of monitoring metrics, respectively. 
Inspired by the factorization machine~\cite{DBLP:conf/icdm/Rendle10/FM}, we reformulate metric and timestamp inner-product pairs to achieve linear time complexity, reducing computational complexity. Both types of features are then concatenated and fed into a cost-effective model for training.



\subsubsection{Multivariate Monitoring Metrics Interactions}

To explicitly capture the dependency between multivariate monitoring metrics and their historical patterns, for each sliding window, denoted as $X_s\in \mathbb{R}^{\omega\times m}$, we calculate the pairwise inner product of all metric feature vectors, i.e., $x_{s\tau,s\tau +\omega - 1}^k, k\in [1, m]$, and temporal vectors, i.e., $x_t, t\in [s\tau, s\tau +\omega - 1]$. Particularly, for notation simplification, we remove the subscript of feature vectors.


\begin{eqnarray}
    h_f &=& b_0 + \sum_{i=1}^mw_ix^i  + \sum_{i=1}^m\sum_{j=i+1}^m \left\langle x^i, x^j\right\rangle v_iv_j\label{equ: dim_oriented} \\
    h_t &=& \hat{b}_0 + \sum_{i=1}^\omega \hat{w}_ix_i  + \sum_{i=1}^\omega\sum_{j=i+1}^\omega \left\langle x_i, x_j \right\rangle \hat{v}_i\hat{v}_j\label{equ: time_oriented}
\end{eqnarray}


The cross-feature and cross-time metric interactions, denoted as $h_f$ and $h_t$, are formulated as Equation~\ref{equ: dim_oriented} and~\ref{equ: time_oriented}, respectively. Particularly, we only elaborate on the Equation~\ref{equ: dim_oriented} as Equation~\ref{equ: time_oriented} shares the same format. In Equation~\ref{equ: dim_oriented}, $b_0, w_i, v_j, v_j\in \mathbb{R}$ are trainable parameters, $x^i$, $x^j \in \mathbb{R}^\omega$ are the $i^{th}$ and $j^{th}$ column of $X_s$ with each column representing all the observations of a metric in the corresponding window, and $<\cdot, \cdot>$ is the operation of inner product. The equation is composed of three terms: the first term is a trainable bias, the second term is a weighted sum of all monitoring metrics without explicit interaction, and the third term is the core part of the proposed collaborative machine, which models the pairwise metric interactions. Specifically, each pair of metric interactions is modeled by the inner product of the corresponding two monitoring metrics. However, not all the pairs share the same importance. To model the importance of each metric and its interaction with other monitoring metrics, we assign a trainable weight to each particular metric, e.g., $v_i$ for the $i^{th}$ metric. After that, while interacting with other monitoring metrics, the product of the weights can measure the importance of the combination, e.g,  $v_iv_j$ denotes the importance of the interaction between the $i^{th}$ and $j^{th}$ monitoring metrics.



\subsubsection{Efficient Computation}

As all pairwise feature and temporal interactions require to be computed, the operations shown in Equation~\ref{equ: dim_oriented} and ~\ref{equ: time_oriented} are computational expensive, i.e., $\mathcal{O}(m^2)$ and $\mathcal{O}(\omega^2)$, respectively. They are not practical for industrial scenarios as a small increase in the number of multivariate monitoring metrics or observation step size, i.e., larger $m$ and $\omega$, will quickly increase the computation complexity. To capture the feature and temporal interactions more efficiently, inspired by the work of Rendle et al.~\cite{DBLP:conf/icdm/Rendle10/FM}, we reformulate the pairwise interaction term in Equation~\ref{equ: dim_oriented} and~\ref{equ: time_oriented}, which could greatly reduce the computation cost, leading to a linear time complexity. Particularly, we only demonstrate the reformulation of the feature interaction term in Equation~\ref{equ: dim_oriented}, as the temporal version in Equation~\ref{equ: time_oriented} can be derived similarly:

\begin{align}
& \sum_{i=1}^{m}\sum_{j=i+1}^{m} \left\langle x^i, x^j  \right\rangle v_iv_j\label{equ: cross_feat_original} \\
=& \frac{1}{2}\sum_{i=1}^{m}\sum_{j=1}^{m} \left\langle x^i, x^j \right\rangle v_iv_j - \frac{1}{2}\sum_{i=1}^{m} \left\langle x^i, x^j \right\rangle v_iv_i\\
=& \frac{1}{2} \left(\sum_{i=1}^{m} \sum_{j=1}^{m} \sum_{r=1}^\omega x^i_rx^j_rv_iv_j - \sum_{i=1}^{m}\sum_{r=1}^\omega x^i_rx^j_rv_iv_i\right)\\
=& \frac{1}{2}\sum^\omega_{r=1}
\left(
\left( \sum_{i=1}^{m} x^i_r v_i \right)
\left( \sum_{j=1}^{m}x^j_rv_j \right) -
\sum_{i=1}^{m} (x^i_r)^2 v_i^2
\right)\\
=& \frac{1}{2}\sum^\omega_{r=1}
\left(                                                  
\left(
\sum_{i=1}^{m} x^i_rv_i
\right)^2 - 
\sum_{i=1}^{m} (x^i_r)^2v_i^2
\right) \numberthis \label{equ: simplified_fm}
\end{align}

Equation~\ref{equ: simplified_fm} is mathematically equivalent to the third term of Equation~\ref{equ: dim_oriented}. 
The time complexity for Equation~\ref{equ: cross_feat_original} is $\mathcal{O}(m^2\omega)$ since the item $\left\langle x^i, x^j  \right\rangle$ needs $\omega$ times computations. After our reformulation, the computation complexity for capturing metric interactions is decreased from $\mathcal{O}(m^2\omega)$ to $\mathcal{O}(m\omega)$ as shown in Equation~\ref{equ: simplified_fm}. Moreover, the computation involved in Equation~\ref{equ: simplified_fm} can be implemented as matrix multiplications, which can be accelerated by GPU (graphics processing unit) for better efficiency. By substituting the interaction terms in both Equation~\ref{equ: dim_oriented} and~\ref{equ: time_oriented}, we can capture the pairwise cross-feature and cross-time interactions to produce $h_f$ and $h_t$ in a highly efficient manner.

\subsection{Model Training and Anomaly Detection}

The last two phases of CMAnomaly are model training and anomaly detection. The anomaly detection model of CMAnomaly works in a forecasting manner. To be more specific, the model is trained to predict the next metric values given preceding observations. Anomaly is then detected based on prediction errors, as shown in Figure~\ref{fig:overall_framework}. In the training phase, as most multivariate monitoring metrics would reflect the normal status of an entity, the model will learn the normal patterns of monitoring metrics, i.e., what the next observations would be given previous ones. Although there could be anomalies in the training data, they tend to be forgotten by the model as they rarely appear. Consequently, in the detection phase, the model will predict "normal" metric values based on the learned patterns. If the real observations deviate from the predicted ones by a significant margin, an anomaly may happen, i.e., the entity is not in its normal status. Particularly, the deviation measures the likelihood of the occurrence of the anomaly.

Our framework supports various types of neural network models for anomaly detection. However, thanks to the formulation of metric interactions in Equation~\ref{equ: dim_oriented} and~\ref{equ: time_oriented}, we find a cost-effective model architecture, i.e., multi-layer perceptron (MLP), is sufficient to achieve competitive performance. The anomaly detection model can be formulated as follows:





\begin{eqnarray}
    \tilde{h}_{i+1} = \sigma(\tilde{h}_i\tilde{X}_i + \tilde{b}_i), i=0,1,..., L-1,
\end{eqnarray}

\noindent where $L$ is the number of layers of the MLP model, $\tilde{W}_i, \tilde{b}_i$ are trainable parameters with customizable size, and $\sigma(x) = max(0, x)$ is the ReLU activation function. Particularly, we simultaneously consider the cross-feature and cross-time metric interactions by concatenating $h_f$ and $h_t$, which is the input to the model, i.e., $\tilde{h}_0 = concat(h_f, h_t)$. $\hat{y} = \tilde{h}_{L} \in \mathbb{R}^m$ is the prediction result produced by the last layer of the MLP model, which contains the predicted values for all monitoring metrics at the next timestamp.



\subsubsection{Training}

We train the anomaly detection model by minimizing the following mean squre error (MSE) loss $\mathcal{L}$ between the predictions and ground truth observations:

\begin{eqnarray}
\mathcal{L} = \sum_{i=1}^N \left\Vert \hat{y}_i - \hat{x}_i \right\Vert_2,
\end{eqnarray}

\noindent where $N$ is the number of training sliding windows. $\hat{y}_i \in \mathbb{R}^m$ and $\hat{x}_i=x_{i\tau +\omega} \in \mathbb{R}^m$ are the predicted and ground truth observations for the $i^{th}$ window, respectively. The minimization of $\mathcal{L}$ is conducted using the Adam optimizer~\cite{DBLP:journals/corr/KingmaB14/adam}. Particularly, we stop the training process when the loss $\mathcal{L}$ is relatively stable, i.e., the change of $\mathcal{L}$ is less than  $10^{-5}$. With the minimization of loss $\mathcal{L}$ during training, CMAnomaly can learn from the normal patterns in the training data by updating all trainable parameters, e.g., $v_iv_j$ denoting the interaction weights. 


\subsubsection{Detection}

After the model is trained, we compute an anomaly score for each window $X_i$ in the testing data. Specifically, we first calculate the MSE between the predicted and ground truth observations, and then apply the sigmoid function to rescale the score to the range $[0, 1]$, which represents the probability of the occurrence of an anomaly:



\begin{eqnarray}
    s_i = \phi\left(\frac{1}{m}  \left\Vert \hat{y}_i - \hat{x}_i \right\Vert_2\right) 
\end{eqnarray}

\noindent where $\phi(x) = \frac {1}{1+e^{-x}}$ is the sigmoid function.

To determine whether an anomaly has happened, a threshold $\theta$ should be defined for the anomaly score. The timestamps with a large anomaly score, i.e., $s_i \geq \theta$, should be regarded as anomalous points. In reality, the threshold can be set by on-site engineers based on their experience. A large threshold imposes a strict anomaly detection policy, which may miss important system failures, i.e., low recall. However, a small threshold increases the sensitivity to metric changes, resulting in false alarms, i.e., low precision. In our experiments, we empirically set the threshold that yields the best experimental results.

\section{Experiments}
\label{sec:experiments}

In this section, we evaluate CMAnomaly using both public data and industrial data. Particularly, we aim to answer the following research questions (RQs):

\textbf{RQ1}: How effective is CMAnomaly in anomaly detection over multivariate monitoring metrics?

\textbf{RQ2}: How much the proposed collaborative machine mechanism contributes to the overall effectiveness of CMAnomaly?

\textbf{RQ3}: How efficient is CMAnomaly compared with existing approaches?


\subsection{Experimental Setup}

\subsubsection{Dataset}

To evaluate the effectiveness of CMAnomaly, we conduct experiments on three public datasets. Moreover, to confirm its practical significance, we collect an industrial dataset from the Networking service of the Huawei Cloud.

\textbf{Public dataset}. 
SMD (Server Machine Dataset). SMAP and MSL are two real-world datasets released by NASA. They are collected from the running spacecraft and contain a set of telemetry anomalies corresponding to actual spacecraft issues involving various subsystems and channel types~\cite{DBLP:conf/kdd/HundmanCLCS18/lstm_ndt}. SMD dataset is collected from a large Internet company containing a 5-week-long monitoring metrics of $28$ machines released by Su et al.~\cite{DBLP:conf/kdd/SuZNLSP19/omni}.

\textbf{Industrial dataset}. Besides the public dataset, we also collected real-world monitoring metrics from a global cloud service provider Huawei Cloud to conduct a more comprehensive evaluation.
The online services of the Huawei Cloud have been supporting tens of millions of users worldwide. Therefore, to provide a stable $24 \times 7$ service, the status of each online service is closely monitored with monitoring metrics.
The engineers can fix problems timely if the anomalies of monitoring metrics can be automatically detected and reported in real-time. To evaluate our method in a practical scenario, we collected a 30-day-long monitoring metrics dataset with 13 online services. 
The dataset is termed as \textit{Industry} in Table~\ref{tab: dataset}. Each of the online services has 15$\sim$30 different types of monitoring metrics. We use the first 20 days of monitoring metrics as the training data and the rest as the testing data.
For labeling the dataset, we rely on corresponding issue reports, which contain the start and end times of problems identified by on-site engineers or customers of an online service. Anomalies are designated during periods when the online service is deemed problematic, while the remaining data is considered normal. 
We also open-source this dataset to facilitate future studies in this field\footnote{https://github.com/OpsPAI/CMAnomaly}.

The statistics of the datasets are listed in the Table~\ref{tab: dataset}. \textit{\#Train} and \textit{\#Test} denote the number of observations in the training and testing dataset, respectively. \textit{\#Entities} denotes the number of entities in a dataset. Specifically, in SMAP and MSL, the entity is a spacecraft that can be viewed as a complex software system composed of multiple interconnected components, each serving a specific functionality. In SMD, the entity is a server machine. In our industrial dataset, each entity serves as an online service, e.g., load balance, machine learning, deep learning, remote storage, and authentication. \textit{\#Dim.} stands for the number of monitoring metrics associated with an entity. \textit{Anomaly Ratio} is the ratio of the number of anomalous points to the number of observations.

\begin{table}[h]
\caption{Dataset Statistics}
\label{tab: dataset}
\begin{tabular}{cccccc}
\hline
Dataset & \#Train & \#Test & \#Entities & \#Dim. & \begin{tabular}[c]{@{}c@{}}Anomaly \\ Ratio (\%)\end{tabular} \\ 
\hline
\hline
SMAP & 135,183 & 427,617 & 55 & 25 & 13.13 \\
MSL & 58,317 & 73,729 & 27 & 55 & 10.72 \\
SMD & 708,405 & 708,420 & 28 & 38 & 4.16 \\
Industry & 1,728,000 & 864,000 & 13 & 15$\sim$30 & 1.16 \\ 
\hline
\end{tabular}
\end{table}

\subsubsection{Experimental Environment}

We run all experiments on a Linux server with Intel Xeon Gold 6148 CPU @ 2.40GHZ and 1TB RAM, GeForce RTX 2080 Ti, running Red Hat 4.8.5 with Linux kernel 3.10.0. In addition, the proposed model is implemented under the PyTorch framework and runs on the GPU. 

\subsubsection{Evaluation Metrics}

As anomaly detection is a binary classification problem, we employ Precision (PC), Recall (RC), and F1 score (F1) for evaluation, as calculated by Equation~\ref{Equ: metrics}.
Specifically, TP (true positive) denotes the number of anomalous samples that the model correctly predicts as an anomaly. FP (false positive) is the number of normal samples that the model mistakenly recognizes as an anomaly. FN (false negative) is the number of normal samples that are predicted as normal. F1 score is the harmonic mean between precision and recall. A higher F1 score indicates a better model. To show the best performance of our method, we manually tune the anomaly threshold to achieve the best F1 score as done by~\cite{ DBLP:conf/kdd/AudibertMGMZ20/usad}.

\small{
\begin{eqnarray}
PC = \frac{TP}{TP + FP} \quad
RC = \frac{TP}{TP + FN} \quad
F1 = 2 \cdot \frac{PC \cdot RC}{PC + RC}\label{Equ: metrics}
\end{eqnarray}
}

In practice, failures often last for a certain period, which results in consecutive anomalies within a time range. 
Therefore, it is practical to apply the \textit{point adjustment} process as done in \cite{DBLP:conf/kdd/SuZNLSP19/omni, DBLP:conf/kdd/AudibertMGMZ20/usad, DBLP:conf/kdd/RenXWYHKXYTZ19/microsoft}, which consider that a model makes a correct prediction for an anomalous segment if at least one point in the segment is successfully predicted as an anomaly. The following evaluation results are obtained after applying point adjustment.



\begin{table*}[t!]
\centering
\caption{Accuracy Comparison of Different Anomaly Detection Methods on Public Datasets}
\includegraphics[width=1.0\textwidth]{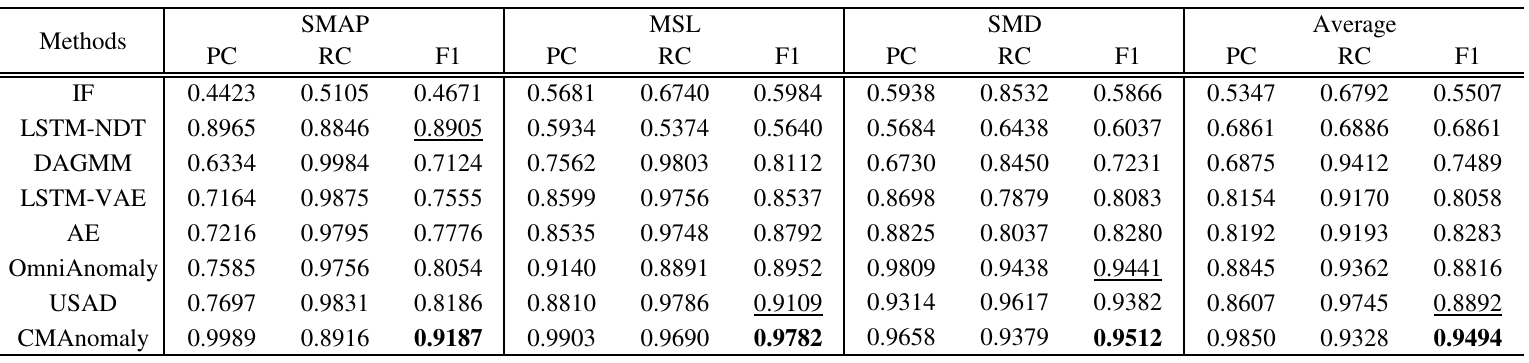}
\label{tab: accuracy}
\vspace{-20pt}
\end{table*}

\subsection{RQ1: Effectiveness of CMAnomaly}

To study the effectiveness of CMAnomaly, we compare its performance with various state-of-the-art baseline models on both the public datasets and the industrial dataset collected from Huawei Cloud. We first train CMAnomaly on the training data until convergence. Then, we compute the anomaly score for each observation of the testing data. After that, we manually try thresholds $\theta$ in the range of $[0,1]$ with a step size of 0.1 to produce the prediction. In particular, for the baselines with automatic threshold selection mechanism, we replace such mechanism with manual selection. In doing this, we follow a recent work~\cite{DBLP:conf/kdd/AudibertMGMZ20/usad} to obtain the best performance of all models for a fair comparison. Finally, the predictions of all models are adjusted to compute PC, RC, and F1.

\textbf{Performance on public datasets}. We select six unsupervised multivariate metric anomaly detection methods as the baselines, which are isolation forest (IF)~\cite{DBLP:conf/icdm/LiuTZ08/isolation_forest}, LSTM-NDT~\cite{DBLP:conf/kdd/HundmanCLCS18/lstm_ndt}, DAGMM~\cite{DBLP:conf/iclr/ZongSMCLCC18/dagmm}, LSTM-VAE~\cite{DBLP:journals/corr/abs-1711-00614/lstm-vae}, autoencoder (AE)~\cite{DBLP:conf/kdd/AudibertMGMZ20/usad}, OmniAnomaly~\cite{DBLP:conf/kdd/SuZNLSP19/omni} and USAD~\cite{DBLP:conf/kdd/AudibertMGMZ20/usad}. 



The overall performance is shown in Table~\ref{tab: accuracy}, where we mark the best F1 score in bold and underline the second-best ones. We also report the average metrics on all datasets in the "Average" columns. We have the following observations:
(1) IF achieves the worst performance compared with other baseline models. Because IF tries to isolate the anomalous timestamp independently, which loses the temporal dependency. 
(2) Other LSTM-based or AE-based methods including LSTM-NDT, DAGMM, LSTM-VAE, AE perform much better than IF and achieve $0.6861 \sim 0.8283$ F1 score because these models take a window of observations as input, which helps to retain valuable historical information. 
(3) OmniAnomaly and USAD outperform other baselines by a large margin and achieve nearly $0.88$ f1 score. These two methods introduce different mechanisms to ensure robust anomaly detection. OmniAnomaly models MTS through stochastic variables and uses reconstruction probabilities to determine anomalies. Differently, USAD utilizes the generative adversarial network (GAN) to train autoencoders, which allows the model to detect anomalies close to normal data.
(4) The proposed method CMAnomaly achieves the highest F1 score on all datasets. The average F1 score is improved to 0.9494 from the second-best, i.e., 0.8892, achieved by USAD. The cost-effective design of CMAnomaly allows it to incur fewer trainable parameters than USAD. As a consequence, overfitting to the training data could be alleviated. In other words, only normal patterns with relatively high occurrence are learned by CMAnomaly during training. When making anomaly detection on the test data, CMAnomaly can be less sensitive and only raise the anomaly score when a real anomaly happens, avoiding more false alarms than USAD. The experimental results indicate that the PC of CMAnomaly is significantly higher than USAD (0.9850 vs 0.8607), which implies fewer false alarms.
 
\begin{table}[h]
\centering
\caption{Accuracy Comparison on Industrial Dataset}
\includegraphics[scale = 0.76]{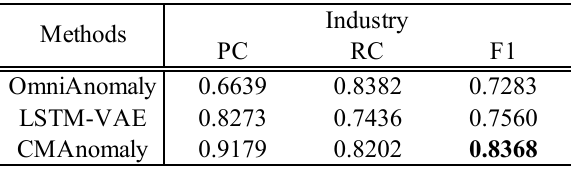}
\label{tab: industry}
\vspace{-10pt}
\end{table}

\textbf{Performance on industrial dataset}. To evaluate the effectiveness of CMAnomaly in the industrial application, we apply two most effective anomaly detection methods whose source code is publicly available, i.e., LSTM-VAE~\cite{DBLP:journals/corr/abs-1711-00614/lstm-vae} and OmniAnomaly~\cite{DBLP:conf/kdd/SuZNLSP19/omni} on the industrial dataset for comparison. 

The experimental results are shown in Table~\ref{tab: industry}. In particular, the PC of OmniAnomaly is the lowest, but the RC is the highest because the complex architecture of OmniAnomaly incurs more trainable parameters, which makes it easier to overfit the training data. Therefore, OmniAnomaly is more sensitive to capture more anomalies but has the most false positive alarms. Different from the results in Table~\ref{tab: accuracy}, LSTM-VAE outperforms OmniAnomaly on this dataset in terms of F1. LSTM-VAE has a more light-weight design than OmniAnomaly, so LSTM-VAE suffers less overfitting. As a result, LSTM-VAE only raises the anomaly score when the new observation deviates more from the prediction. In this case, though higher PC is achieved, LSTM-VAE has the lowest RC because it cannot effectively find all possible anomalies.
CMAnomaly can balance PC and RC better and achieves the best F1 score, $\sim$0.08 higher than the second-best one achieved by LSTM-VAE. The collaborative machine facilitates CMAnomaly to capture the dependency of the training monitoring metrics effectively. Therefore, CMAnomaly avoids overfitting the noisy points existing in training data e.g., usual spikes as shown in Figure~\ref{fig: dependency}. Instead, CMAnomaly reports a higher anomaly score only when the dependent monitoring metrics are anomalous, thus achieves the highest precision. Moreover, CMAnomaly keeps the sensitivity to detect more true positive samples thanks to the ability of capturing the dependency.


\subsection{RQ2: Effectiveness of Collaborative Machine}
\begin{figure*}[ht]
\centering    
\includegraphics[width=1.8\columnwidth]{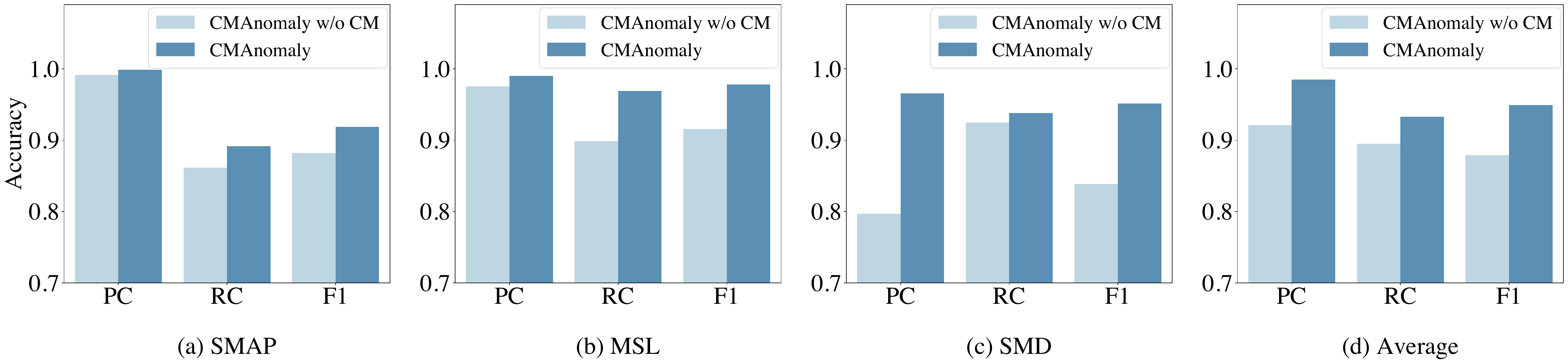}
\vspace{-10pt}
\caption{Performance Comparison with and without Collaborative Machine} 
\label{fig:ablation}  
\vspace{-16pt}
\end{figure*}

To evaluate the effectiveness of the proposed collaborative machine used in CMAnomaly. We firstly produce a variant of CMAnomaly by removing the collaborative machine (CM). Specifically, we use the average of all the monitoring metrics across the time dimension as the input of the MLP instead of hidden vectors computed by Equation~\ref{equ: dim_oriented} and Equation~\ref{equ: time_oriented}. In doing this, explicit interactions are removed. Then we apply the variant to perform anomaly detection on three public datasets SMAP, MSL and SMD. Then, we compare the F1 score obtained by the variant with CMAnomaly.
Figure~\ref{fig:ablation} shows the accuracy comparison between the variant and CMAnomaly. The variant is denoted as "CMAnomaly w/o CM" in the figure. We can observe that the accuracy of the variant drops on all three datasets. Especially, the average accuracy decreases by $\sim 0.07$ from $0.9494$ to $0.8789$. This can be explained by the loss of valuable temporal and feature information of the variant. On the one hand, without the collaborative machine, the sequential dependency and feature dependency cannot be captured,  which hinders the model to learn from the normal patterns in the training data. As a consequence, when making the prediction in the anomaly detection stage, the new normal observation cannot be accurately predicted thus larger anomaly score may be produced. Therefore, more false alarms are reported.

\subsection{RQ3: Efficiency of CMAnomaly}

In this section, we evaluate the efficiency of CMAnomaly compared with the baseline models. We select the OmniAnomaly and LSTM-VAE for comparison because they are the most effective methods that are open-source. We conduct the comparison based on the official implementation of OmniAnomaly\footnote{https://github.com/NetManAIOps/OmniAnomaly} and LSTM-VAE\footnote{https://github.com/TimyadNyda/Variational-Lstm-Autoencoder}.
Next, we perform these methods on the industrial dataset and record training time and prediction time. Training time defines as the total time of feeding the preprocessed data to the model, and prediction is the total time used to predict the next observation on all the windows in the test set. Specifically, for a fair comparison, all the models are trained in a batch-wise manner for one epoch, and the same batch size is used.
In real-world practice, we need to apply different window sizes to detect anomalies under different granularities. Therefore we increase the window size $\omega$ in the range of $[16, 32, 64, 128, 256]$ to evaluate the efficiency of different methods. Stride is set as $\tau=5$ on the training data and $\tau=1$ on the test data because anomaly detection on test data should be conducted under finer granularity to alleviate missing anomalies.
All the experiments run on the same computation machine on a GPU of GeForce RTX 2080 Ti.

\begin{figure}[t]
  \centering
  \mbox{
     \subfigure[Training Time v.s. Window Size\label{fig: efficiency_train}]{\includegraphics[width=0.49\columnwidth]{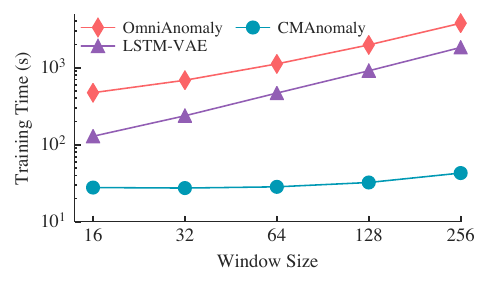}}\quad
    
    \subfigure[Prediction Time vs Window Size
    \label{fig: efficiency_test}]
    {\includegraphics[width=0.49\columnwidth]{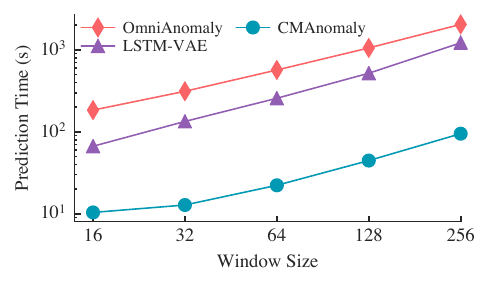}}
    }
   \vspace{-12pt}
  \caption{Efficiency of \nm}
\vspace{-16pt}
\end{figure}

Figure~\ref{fig: efficiency_train} and ~\ref{fig: efficiency_test} show the efficiency comparison of the selected models in the training and prediction phases, respectively. According to the results, we can see that 
(1) the computational cost of LSTM-VAE and OmniAnomaly increases quickly with a larger window size, because both of the methods adopt LSTM as a core part to handle sequential input. Therefore, when we increase the window size, the depth of the LSTM increases, thus incurs more computation.
(2) OmniAnomaly costs more time (nearly $2 \times$) than LSTM-VAE though they share similar LSTM and VAE architecture. Nevertheless, OmniAnomaly introduces additional stochastic variable connection and planar normalizing flow to model the monitoring metrics, which are more computationally expensive.
(3) CMAnomaly is the most efficient method. To be more specific, CMAnomaly is nearly 10x faster than LSTM-VAE and 20x than OmniAnomaly. Significantly, the training and prediction time do not increase a lot when varying the window size. Mainly, since $\tau$ on the test set is set to $1$ and less than that of the training data, more test sliding windows are generated. Therefore, the test time of CMAnomaly is slightly larger than the training time. This only happens on CMAnomaly because the other two methods need more computation on the back-propagation during training.
The cost-effective design and reduction of the computation complexity of CMAnomaly benefit its efficiency. Moreover, both the collaborative machine and MLP of CMAnomaly can be well accelerated by the asynchronous computing architecture of GPUs. Therefore, CMAnomaly could conduct efficient training and prediction.

We want to emphasize that the efficiency of CMAnomaly can meet the demanding requirements of industrial scenarios. (1) Generally, a large-scale system is closely monitored by a large number of monitoring metrics collected every second. 
Such a large quantity of monitoring metrics cloud be processed by CMAnomaly in real-time. (2) CMAnomaly can be fast retrained to learn the new patterns of monitoring metrics after the system upgrades. With this advantage, the deployment of an online anomaly detection service will not be suspended for too long.


\subsection{Case Study}

\begin{figure}[h]
\centering
\includegraphics[width=0.8\columnwidth]{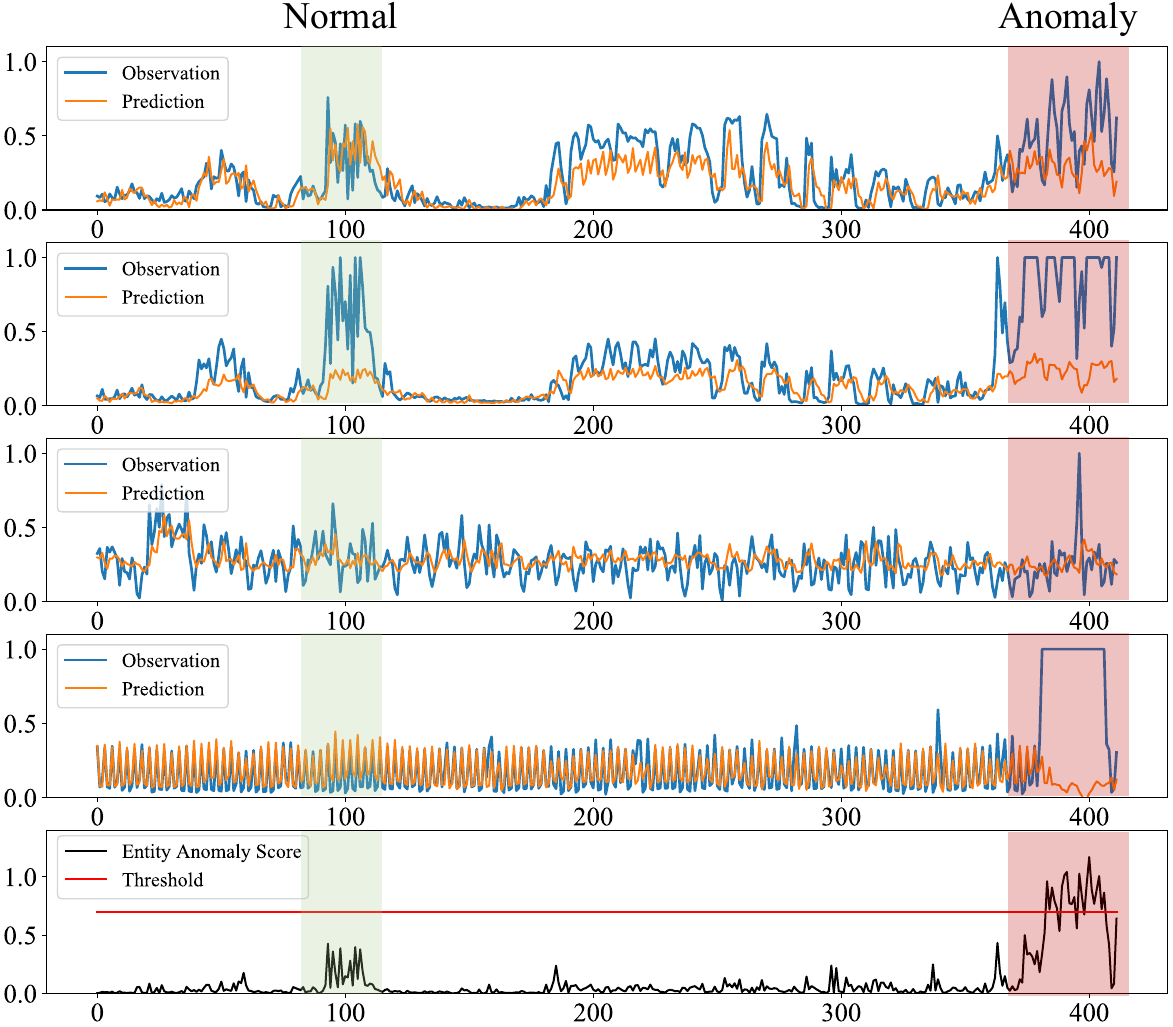}
\vspace{-10pt}
\caption{Case Study on Industrial Dataset}
\label{fig: case_study}
\vspace{-18pt}
\end{figure}

To make a more concrete evaluation of CMAnomaly, we provide a case study on how CMAnomaly computes anomaly scores for given multivariate monitoring metrics. This case study is conducted on a segment of the industrial dataset. The monitoring metrics monitor the traffic of connected network nodes supporting a service. After the model is trained, we apply the model to consecutively predict the monitoring metrics and detect anomalies. The results are shown in Figure~\ref{fig: case_study}. In this figure, the first four rows show the real observation of the monitoring metrics and the prediction made by CMAnomaly. The last row shows the anomaly score of the entity and a threshold selected. If we consider the monitoring metrics individually, the first two monitoring metrics have spikes at both segments marked as "Normal" and "Anomaly" so anomalies would be alarmed. Differently, our method computes a much lower anomaly score of the "Normal" segment than the "Anomaly" segment by simultaneously considering the dependency of all the monitoring metrics. As a result, the anomaly score at the true positive segment can be magnified. In fact, the "Normal" spikes are caused by a typical increase of user requests to a port of the service. However, in the "Anomaly" segment, a physical machine failure happens, which leads to anomalous patterns of the monitoring metrics.

\subsection{Threat to Validity}

We have identified the following major threats to validity.

\textbf{Labeling noise}. To evaluate the practical usage of CMAnomaly, we conduct experiments on an industrial dataset collected from \cpn. The evaluation requires on-site engineers to manually inspect the monitoring metrics and label the anomalies. As the labeling is done manually, label noises, i.e., false positives and false negatives, may be introduced. However, the engineers we invited have rich domain knowledge and years of system troubleshooting experience. They are required to discuss and confirm the labeling results with each other. Moreover, the failures we selected are typical to the service systems. Engineers can quickly and confidently do the labeling work. Therefore, we believe the amount of noise is small.

\textbf{Model implementation and parameter settings}. The implementation of our model and baseline models as well as parameter settings are important internal threats to the validity. To reduce this threat, we collect the source code released by each work or directly employ the results employed in the paper. For our model, we employ peer code review. Specifically, all authors of this paper are required to check the code carefully. Moreover, all experiments are repeated multiple times and the average results are reported. In terms of parameter settings, we conduct a series of comparative experiments and empirically select the best results. As our model possesses the merit of high efficiency, the best results can be quickly obtained.
\section{Related Work}
\label{sec:related_work}
In the literature, anomaly detectors over multivariate metrics can be categorized to \textit{implicit modeling} and \textit{explicit modeling} schemas.

\textbf{Implicit Modeling.} The major challenge of multivariate anomaly detection comes from the effective modeling of complex temporal dependence and stochasticity of Multivariate monitoring metrics. 
Much previous work attempts to capture the normal temporal dependence by modeling hidden states implicitly.
For example, Hundman et al.~\cite{DBLP:conf/kdd/HundmanCLCS18/lstm_ndt} leverages LSTM for anomaly detection in spacecraft time-series metrics based on prediction errors, 
Malhotra et al. \cite{DBLP:journals/corr/MalhotraRAVAS16} proposes an LSTM-based encoder-decoder network that reconstructs ``normal" time series and detects anomalies through reconstruction errors.
Another way to model normal patterns is to learn the distribution of input data like deep generative models~\cite{DBLP:conf/nips/GoodfellowPMXWOCB14} and deep Bayesian network~\cite{DBLP:journals/tkde/WangY16}. 
Donut~\cite{DBLP:conf/www/XuCZLBLLZPFCWQ18/donut} employs Variational AutoEncoder (VAE) to generate the normal hidden state of seasonal monitoring metrics without expert-labeled data.
With solid theoretical explanation, Donut successfully detects anomalies in seasonal monitoring metrics with various patterns and data quality, but enjoys high time complexity in the training phase.
To reduce the training complexity, DAGMM~\cite{DBLP:conf/iclr/ZongSMCLCC18/dagmm} simplify the hidden state as a combination of several Gaussian distributions.
USAD~\cite{DBLP:conf/kdd/AudibertMGMZ20/usad} improves the autoencoder framework by incorporating adversarial samples to speed up the training phase.
OmniAnomaly~\cite{DBLP:conf/kdd/SuZNLSP19/omni} uses a stochastic recurrent neural network to simulate the normal distribution of multivariate time-series data. Anomalies are identified when the input data reconstruction has low probability. 
However, the generalization capability of these generative approaches with implicitly modeling is degraded when they encounter severe noise in temporal monitoring metrics, which is very common in industrial production systems. 

\textbf{Explicit Modeling.} To reduce the negative effect of noisy data in multivariate monitoring metrics, the temporal dependencies such as the inter-correlations between different parts of monitoring metrics should be captured in an explicit way. 
Zhang et al.~\cite{DBLP:conf/aaai/ZhangSCFLCNZCC19} proposes a convolutional recurrent encoder-decoder framework called MSCRED to characterize multiple levels of the normal states in different steps of monitoring metrics.
MSCRED employs a convolutional encoder and convolutional LSTM network to capture the monitoring metric interactions and temporal patterns, respectively.
Finally, it generates different levels of anomaly scores based upon the severity of different incidents.
Similar to our approach, Zhao et al.~\cite{DBLP:conf/icdm/ZhaoWDHCTXBTZ20} consider each univariate monitoring metric as an individual feature to capture the complex dependencies of multivariate monitoring metrics from temporal and feature perspective.
Their proposed method incorporates feature-oriented and time-oriented graph attention mechanisms to obtain mixed hidden representation from a combination of forecasting-based and reconstruction-based frameworks.
However, the overall framework enjoys high time-complexity and is hard to address the demanding industrial requirements for quick response and adjustment.

\section{Industrial Experience}
\label{sec:discussions}

\subsection{Success Story}

In Huawei Cloud, \nm~has been successfully incorporated into the troubleshooting system for two years, serving several essential services that have a large number of customers worldwide. We conduct field interviews with on-site engineers to collect feedback. Based on the feedback, we have seen it shedding light on highly accurate and efficient anomaly detection on multivariate monitoring metrics. Due to the complex system architecture, in Huawei Cloud, the number of monitoring metrics being monitored is extremely large. Existing anomaly detection algorithms either report too many false positives or cannot meet the efficiency requirements, i.e., real-time detection and fast model retraining. Owing to the proposed mechanism of collaborative machines and cost-effective model architecture, \nm~ can achieve superior performance.

\subsection{Lessons Learned}
\textbf{Model Maintenance}. Monitoring the accuracy of a deployed model is crucial in its industrial application. It is common that the accuracy of a model decreases over time due to the notorious \textit{concept drift} problem~\cite{chen2022adaptive}. 
In our scenario, the accuracy drop can usually happen when a new version of an online service is launched, which incurs different anomaly patterns in the monitoring metrics. An effective solution to this problem is to retrain the anomaly detection model with newly collected data. At Huawei Cloud, a substantial number of monitoring metrics are generated on a daily basis. Consequently, it becomes necessary for the model to be retrained within a short period of time, allowing for the rapid redeployment of a new version. Thanks to the lightweight design of CMAnomaly, we are able to establish an efficient pipeline for updating the model. This pipeline encompasses various stages, including data collection, retraining, evaluation, and deployment.

\textbf{Utilizing Expert Knowledge}. On-site engineers often possess decent knowledge about the systems. Such expert knowledge can facilitate model design and development. For example, given thousands of monitoring metrics at hand, they can quickly select the important ones that best characterize the systems' health status. Moreover, instead of finding dependent monitoring metrics from scratch as done in this paper, engineers can help identify a small set of correlated monitoring metrics first. A more accurate model can then be pursued on top of these monitoring metrics. However, such valuable knowledge is often not well accumulated, organized, and documented. For incident management in cloud systems, automated knowledge extraction~\cite{DBLP:journals/corr/abs-2007-05505} has been proposed. Efforts should also be devoted to system troubleshooting based on monitoring metrics.

\textbf{Building data collection pipeline}. For modern software systems, IT operations play a crucial role for system reliability assurance~\cite{DBLP:conf/icse/DangLH19,DBLP:conf/icse/0003HLXZHGXDZ19,DBLP:conf/sigsoft/ZhaoCWPWWZFNZSP20,DBLP:conf/icse/Zhao0PWWZCZNWWZ20,DBLP:conf/sigsoft/LinHDZSXLLWYCZ18}. Since it is data-driven by nature, sufficient and high-quality data are the foundation for model development in many applications, e.g., anomaly detection, failure diagnosis, fault localization. However, common data quality issues include insufficient labels, data noise, etc. To alleviate this issue, a complete and efficient pipeline should be constructed for monitoring data (e.g., monitoring metrics and logs~\cite{DBLP:conf/kbse/HeCHL18,DBLP:conf/sosp/XuHFPJ09,DBLP:conf/icse/LinZLZC16}) collection and storage. Specifically, when failure is detected, there should be tools to facilitate the labeling work for engineers. The labeled data should also be properly stored in a database which supports easy query in future. Once there are sufficient labels, possible supervised learning models can be explored.

\section{Conclusion}
\label{sec:conclusion}

In this paper, we propose CMAnomaly, an anomaly detection framework for multivariate monitoring metrics based on collaborative machine. Particularly, the proposed collaborative machine can learn the pairwise cross-feature and cross-time interactions between monitoring metrics with linear time complexity. Thus, CMAnomaly can quickly obtain a big picture of a system's health status for anomaly detection. We have conducted experiments with three public datasets and one industrial dataset collected from a large online service of Huawei Cloud. For public datasets, CMAnomaly has achieved an average F1 score of 0.9494, outperforming the existing best approach by 6.77\%. Similarly, a 10.68\% of accuracy gain is achieved in the industrial dataset. More importantly, for model training and prediction, CMAnomaly runs up to 20x faster than the baseline methods, demonstrating that CMAnomaly is capable of meeting the demanding industrial requirements in terms of effectiveness and efficiency. Our framework has been successfully incorporated into the troubleshooting system of the \cpn. Feedback from on-site engineers confirms its practical usefulness.



\section{Acknowledgement}
The work described in this paper was supported by the Key-Area Research and Development Program of Guangdong Province (No. 2020B010165002) and the Key Program of Fundamental Research from Shenzhen Science and Technology Innovation Commission (No. JCYJ20200109113403826). It was also supported by the Research Grants Council of the Hong Kong Special Administrative Region, China (No. CUHK 14206921 of the General Research Fund) and the National Natural Science Foundation of China (No. 62202511).

\balance
\bibliographystyle{IEEEtran}
\bibliography{issre23}

\begin{thebibliography}{10}
\providecommand{\url}[1]{#1}
\csname url@samestyle\endcsname
\providecommand{\newblock}{\relax}
\providecommand{\bibinfo}[2]{#2}
\providecommand{\BIBentrySTDinterwordspacing}{\spaceskip=0pt\relax}
\providecommand{\BIBentryALTinterwordstretchfactor}{4}
\providecommand{\BIBentryALTinterwordspacing}{\spaceskip=\fontdimen2\font plus
\BIBentryALTinterwordstretchfactor\fontdimen3\font minus
  \fontdimen4\font\relax}
\providecommand{\BIBforeignlanguage}[2]{{%
\expandafter\ifx\csname l@#1\endcsname\relax
\typeout{** WARNING: IEEEtran.bst: No hyphenation pattern has been}%
\typeout{** loaded for the language `#1'. Using the pattern for}%
\typeout{** the default language instead.}%
\else
\language=\csname l@#1\endcsname
\fi
#2}}
\providecommand{\BIBdecl}{\relax}
\BIBdecl

\bibitem{DBLP:conf/sigsoft/ChenKLZZXZYSXDG20}
Z.~Chen, Y.~Kang, L.~Li, X.~Zhang, H.~Zhang, H.~Xu, Y.~Zhou, L.~Yang, J.~Sun,
  Z.~Xu, Y.~Dang, F.~Gao, P.~Zhao, B.~Qiao, Q.~Lin, D.~Zhang, and M.~R. Lyu,
  ``Towards intelligent incident management: why we need it and how we make
  it,'' in \emph{Proceedings of the 28th Joint European Software Engineering
  Conference and Symposium on the Foundations of Software Engineering,
  (ESEC/FSE)}, 2020, pp. 1487--1497.

\bibitem{DBLP:conf/usenix/ZhangLXQ0QDYCCW19}
X.~Zhang, Q.~Lin, Y.~Xu, S.~Qin, H.~Zhang, B.~Qiao, Y.~Dang, X.~Yang, Q.~Cheng,
  M.~Chintalapati, Y.~Wu, K.~Hsieh, K.~Sui, X.~Meng, Y.~Xu, W.~Zhang, F.~Shen,
  and D.~Zhang, ``Cross-dataset time series anomaly detection for cloud
  systems,'' in \emph{Proceedings of the Annual Technical Conference, (USENIX
  ATC)}, 2019, pp. 1063--1076.

\bibitem{DBLP:conf/icse/DangLH19}
Y.~Dang, Q.~Lin, and P.~Huang, ``Aiops: real-world challenges and research
  innovations,'' in \emph{Proceedings of the 41st International Conference on
  Software Engineering: Companion Proceedings, (ICSE)}, 2019, pp. 4--5.

\bibitem{DBLP:journals/corr/abs-2009-07237}
S.~He, P.~He, Z.~Chen, T.~Yang, Y.~Su, and M.~R. Lyu, ``A survey on automated
  log analysis for reliability engineering,'' \emph{CoRR}, vol. abs/2009.07237,
  2020.

\bibitem{DBLP:conf/sigsoft/HeLLZLZ18}
S.~He, Q.~Lin, J.~Lou, H.~Zhang, M.~R. Lyu, and D.~Zhang, ``Identifying
  impactful service system problems via log analysis,'' in \emph{Proceedings of
  the Joint Meeting on European Software Engineering Conference and Symposium
  on the Foundations of Software Engineering, (ESEC/FSE) {FSE}}.\hskip 1em plus
  0.5em minus 0.4em\relax {ACM}, 2018, pp. 60--70.

\bibitem{DBLP:journals/corr/abs-2302-05092}
\BIBentryALTinterwordspacing
C.~Lee, T.~Yang, Z.~Chen, Y.~Su, and M.~R. Lyu, ``Eadro: An end-to-end
  troubleshooting framework for microservices on multi-source data,''
  \emph{CoRR}, vol. abs/2302.05092, 2023. [Online]. Available:
  \url{https://doi.org/10.48550/arXiv.2302.05092}
\BIBentrySTDinterwordspacing

\bibitem{chen2022adaptive}
Z.~Chen, J.~Liu, Y.~Su, H.~Zhang, X.~Ling, Y.~Yang, and M.~R. Lyu, ``Adaptive
  performance anomaly detection for online service systems via pattern
  sketching,'' in \emph{Proceedings of the 44th International Conference on
  Software Engineering}, 2022, pp. 61--72.

\bibitem{liu2023incident}
J.~Liu, S.~He, Z.~Chen, L.~Li, Y.~Kang, X.~Zhang, P.~He, H.~Zhang, Q.~Lin,
  Z.~Xu \emph{et~al.}, ``Incident-aware duplicate ticket aggregation for cloud
  systems,'' \emph{arXiv preprint arXiv:2302.09520}, 2023.

\bibitem{liu2023prism}
J.~Liu, Z.~Jiang, J.~Gu, J.~Huang, Z.~Chen, C.~Feng, Z.~Yang, Y.~Yang, and
  M.~R. Lyu, ``Prism: Revealing hidden functional clusters from massive
  instances in cloud systems,'' \emph{arXiv preprint arXiv:2308.07638}, 2023.

\bibitem{li2022intelligent}
Y.~Li, X.~Zhang, S.~He, Z.~Chen, Y.~Kang, J.~Liu, L.~Li, Y.~Dang, F.~Gao, Z.~Xu
  \emph{et~al.}, ``An intelligent framework for timely, accurate, and
  comprehensive cloud incident detection,'' \emph{ACM SIGOPS Operating Systems
  Review}, vol.~56, no.~1, pp. 1--7, 2022.

\bibitem{DBLP:conf/kdd/RenXWYHKXYTZ19/microsoft}
H.~Ren, B.~Xu, Y.~Wang, C.~Yi, C.~Huang, X.~Kou, T.~Xing, M.~Yang, J.~Tong, and
  Q.~Zhang, ``Time-series anomaly detection service at microsoft,'' in
  \emph{Proceedings of the 25th International Conference on Knowledge Discovery
  {\&} Data Mining, (KDD)}, 2019, pp. 3009--3017.

\bibitem{DBLP:conf/www/XuCZLBLLZPFCWQ18/donut}
H.~Xu, W.~Chen, N.~Zhao, Z.~Li, J.~Bu, Z.~Li, Y.~Liu, Y.~Zhao, D.~Pei, Y.~Feng,
  J.~Chen, Z.~Wang, and H.~Qiao, ``Unsupervised anomaly detection via
  variational auto-encoder for seasonal kpis in web applications,'' in
  \emph{Proceedings of the 2018 World Wide Web Conference on World Wide Web,
  (WWW)}.\hskip 1em plus 0.5em minus 0.4em\relax {ACM}, 2018, pp. 187--196.

\bibitem{DBLP:journals/corr/abs-2004-00433/univariate_survey}
M.~Braei and S.~Wagner, ``Anomaly detection in univariate time-series: {A}
  survey on the state-of-the-art,'' \emph{CoRR}, vol. abs/2004.00433, 2020.

\bibitem{DBLP:journals/tkde/GuptaGAH14/anomaly_survey2}
M.~Gupta, J.~Gao, C.~C. Aggarwal, and J.~Han, ``Outlier detection for temporal
  data: {A} survey,'' \emph{{IEEE} Trans. Knowl. Data Eng.}, vol.~26, no.~9,
  pp. 2250--2267, 2014.

\bibitem{DBLP:conf/kdd/SuZNLSP19/omni}
Y.~Su, Y.~Zhao, C.~Niu, R.~Liu, W.~Sun, and D.~Pei, ``Robust anomaly detection
  for multivariate time series through stochastic recurrent neural network,''
  in \emph{Proceedings of the 25th International Conference on Knowledge
  Discovery {\&} Data Mining, (KDD)}, 2019, pp. 2828--2837.

\bibitem{DBLP:journals/corr/MalhotraRAVAS16}
P.~Malhotra, A.~Ramakrishnan, G.~Anand, L.~Vig, P.~Agarwal, and G.~Shroff,
  ``Lstm-based encoder-decoder for multi-sensor anomaly detection,''
  \emph{CoRR}, vol. abs/1607.00148, 2016.

\bibitem{DBLP:conf/aaai/ZhangSCFLCNZCC19}
C.~Zhang, D.~Song, Y.~Chen, X.~Feng, C.~Lumezanu, W.~Cheng, J.~Ni, B.~Zong,
  H.~Chen, and N.~V. Chawla, ``A deep neural network for unsupervised anomaly
  detection and diagnosis in multivariate time series data,'' in
  \emph{Proceedings of the 33rd Applications of Artificial Intelligence
  Conference, (AAAI)}, 2019, pp. 1409--1416.

\bibitem{DBLP:journals/corr/abs-1711-00614/lstm-vae}
D.~Park, Y.~Hoshi, and C.~C. Kemp, ``A multimodal anomaly detector for
  robot-assisted feeding using an lstm-based variational autoencoder,''
  \emph{CoRR}, vol. abs/1711.00614, 2017.

\bibitem{DBLP:conf/icdm/ZhaoWDHCTXBTZ20}
H.~Zhao, Y.~Wang, J.~Duan, C.~Huang, D.~Cao, Y.~Tong, B.~Xu, J.~Bai, J.~Tong,
  and Q.~Zhang, ``Multivariate time-series anomaly detection via graph
  attention network,'' in \emph{Proceedings of the 20th International
  Conference on Data Mining, (ICDM)}, 2020, pp. 841--850.

\bibitem{DBLP:journals/tnn/ScarselliGTHM09}
F.~Scarselli, M.~Gori, A.~C. Tsoi, M.~Hagenbuchner, and G.~Monfardini, ``The
  graph neural network model,'' \emph{{IEEE} Trans. Neural Networks}, vol.~20,
  no.~1, pp. 61--80, 2009.

\bibitem{DBLP:conf/icdm/Rendle10/FM}
S.~Rendle, ``Factorization machines,'' in \emph{Proceedings of the 10th
  International Conference on Data Mining, (ICDM)}, 2010, pp. 995--1000.

\bibitem{DBLP:conf/kdd/HundmanCLCS18/lstm_ndt}
K.~Hundman, V.~Constantinou, C.~Laporte, I.~Colwell, and
  T.~S{\"{o}}derstr{\"{o}}m, ``Detecting spacecraft anomalies using lstms and
  nonparametric dynamic thresholding,'' in \emph{Proceedings of the 24th
  International Conference on Knowledge Discovery {\&} Data Mining, (KDD)},
  2018, pp. 387--395.

\bibitem{DBLP:conf/ccs/Du0ZS17/deeplog}
M.~Du, F.~Li, G.~Zheng, and V.~Srikumar, ``Deeplog: Anomaly detection and
  diagnosis from system logs through deep learning,'' in \emph{Proceedings of
  the 2017 Conference on Computer and Communications Security, (CCS)}.\hskip
  1em plus 0.5em minus 0.4em\relax {ACM}, 2017, pp. 1285--1298.

\bibitem{DBLP:journals/corr/KingmaB14/adam}
D.~P. Kingma and J.~Ba, ``Adam: {A} method for stochastic optimization,'' in
  \emph{Proceedings of 3rd International Conference on Learning
  Representations, (ICLR)}, 2015.

\bibitem{DBLP:conf/kdd/AudibertMGMZ20/usad}
J.~Audibert, P.~Michiardi, F.~Guyard, S.~Marti, and M.~A. Zuluaga, ``{USAD:}
  unsupervised anomaly detection on multivariate time series,'' in
  \emph{Proceedings of the 26th {SIGKDD} Conference on Knowledge Discovery and
  Data Mining, (KDD)}.\hskip 1em plus 0.5em minus 0.4em\relax {ACM}, 2020, pp.
  3395--3404.

\bibitem{DBLP:conf/icdm/LiuTZ08/isolation_forest}
F.~T. Liu, K.~M. Ting, and Z.~Zhou, ``Isolation forest,'' in \emph{Proceedings
  of the 8th International Conference on Data Mining (ICDM)}, 2008, pp.
  413--422.

\bibitem{DBLP:conf/iclr/ZongSMCLCC18/dagmm}
B.~Zong, Q.~Song, M.~R. Min, W.~Cheng, C.~Lumezanu, D.~Cho, and H.~Chen, ``Deep
  autoencoding gaussian mixture model for unsupervised anomaly detection,'' in
  \emph{Proceedings of the 6th International Conference on Learning
  Representations, (ICLR)}, 2018.

\bibitem{DBLP:conf/nips/GoodfellowPMXWOCB14}
I.~J. Goodfellow, J.~Pouget{-}Abadie, M.~Mirza, B.~Xu, D.~Warde{-}Farley,
  S.~Ozair, A.~C. Courville, and Y.~Bengio, ``Generative adversarial nets,'' in
  \emph{Proceedings of the 27th Conference on Neural Information Processing
  Systems 2014, (NeurIPS)}, 2014, pp. 2672--2680.

\bibitem{DBLP:journals/tkde/WangY16}
H.~Wang and D.~Yeung, ``Towards bayesian deep learning: {A} framework and some
  existing methods,'' \emph{{IEEE} Trans. Knowl. Data Eng.}, vol.~28, no.~12,
  pp. 3395--3408, 2016.

\bibitem{DBLP:journals/corr/abs-2007-05505}
M.~Shetty, C.~Bansal, S.~Kumar, N.~Rao, N.~Nagappan, and T.~Zimmermann,
  ``Neural knowledge extraction from cloud service incidents,'' \emph{CoRR},
  vol. abs/2007.05505, 2020.

\bibitem{DBLP:conf/icse/0003HLXZHGXDZ19}
J.~Chen, X.~He, Q.~Lin, Y.~Xu, H.~Zhang, D.~Hao, F.~Gao, Z.~Xu, Y.~Dang, and
  D.~Zhang, ``An empirical investigation of incident triage for online service
  systems,'' in \emph{Proceedings of the 41st International Conference on
  Software Engineering: Software Engineering in Practice, (ICSE-SEIP)}, 2019,
  pp. 111--120.

\bibitem{DBLP:conf/sigsoft/ZhaoCWPWWZFNZSP20}
N.~Zhao, J.~Chen, Z.~Wang, X.~Peng, G.~Wang, Y.~Wu, F.~Zhou, Z.~Feng, X.~Nie,
  W.~Zhang, K.~Sui, and D.~Pei, ``Real-time incident prediction for online
  service systems,'' in \emph{Proceedings of the 28th Joint Meeting on European
  Software Engineering Conference and Symposium on the Foundations of Software
  Engineering, (ESEC/FSE)}.\hskip 1em plus 0.5em minus 0.4em\relax {ACM}, 2020,
  pp. 315--326.

\bibitem{DBLP:conf/icse/Zhao0PWWZCZNWWZ20}
N.~Zhao, J.~Chen, X.~Peng, H.~Wang, X.~Wu, Y.~Zhang, Z.~Chen, X.~Zheng, X.~Nie,
  G.~Wang, Y.~Wu, F.~Zhou, W.~Zhang, K.~Sui, and D.~Pei, ``Understanding and
  handling alert storm for online service systems,'' in \emph{Proceedings of
  the 42nd International Conference on Software Engineering, Software
  Engineering in Practice, (ICSE-SEIP)}, 2020, pp. 162--171.

\bibitem{DBLP:conf/sigsoft/LinHDZSXLLWYCZ18}
Q.~Lin, K.~Hsieh, Y.~Dang, H.~Zhang, K.~Sui, Y.~Xu, J.~Lou, C.~Li, Y.~Wu,
  R.~Yao, M.~Chintalapati, and D.~Zhang, ``Predicting node failure in cloud
  service systems,'' in \emph{Proceedings of the Joint Meeting on European
  Software Engineering Conference and Symposium on the Foundations of Software
  Engineering, (ESEC/FSE)}, 2018, pp. 480--490.

\bibitem{DBLP:conf/kbse/HeCHL18}
P.~He, Z.~Chen, S.~He, and M.~R. Lyu, ``Characterizing the natural language
  descriptions in software logging statements,'' in \emph{Proceedings of the
  33rd International Conference on Automated Software Engineering, (ASE)},
  2018, pp. 178--189.

\bibitem{DBLP:conf/sosp/XuHFPJ09}
W.~Xu, L.~Huang, A.~Fox, D.~A. Patterson, and M.~I. Jordan, ``Detecting
  large-scale system problems by mining console logs,'' in \emph{Proceedings of
  the 22nd Symposium on Operating Systems Principles, (SOSP)}, 2009, pp.
  117--132.

\bibitem{DBLP:conf/icse/LinZLZC16}
Q.~Lin, H.~Zhang, J.~Lou, Y.~Zhang, and X.~Chen, ``Log clustering based problem
  identification for online service systems,'' in \emph{Proceedings of the 38th
  International Conference on Software Engineering, (ICSE)}, 2016, pp.
  102--111.

\end{thebibliography}
\balance
\end{document}